\documentclass[preprint,showpacs,preprintnumbers,amsmath,amssymb]{revtex4}

\usepackage{graphicx}
\usepackage{dcolumn}
\usepackage{bm}

\begin{document}
\newcommand{\Arg}[1]{\mbox{Arg}\left[#1\right]}
\newcommand{\bb}{\mathbf}
\newcommand{\braopket}[3]{\left \langle #1\right| \hat #2 \left|#3 \right \rangle}
\newcommand{\braket}[2]{\langle #1|#2\rangle}
\newcommand{\be}{\[}
\newcommand{\br}{\vspace{4mm}}
\newcommand{\bra}[1]{\langle #1|}
\newcommand{\braketbraket}[4]{\langle #1|#2\rangle\langle #3|#4\rangle}
\newcommand{\braop}[2]{\langle #1| \hat #2}
\newcommand{\dd}[1]{ \! \! \!  \mbox{d}#1\ }
\newcommand{\DD}[2]{\frac{\! \! \! \mbox d}{\mbox d #1}#2}
\renewcommand{\det}[1]{\mbox{det}\left(#1\right)}
\newcommand{\ee}{\]} 
\newcommand{\eg}{\textbf{\\  Example: \ \ \ }}
\newcommand{\Imag}[1]{\mbox{Im}\left(#1\right)}
\newcommand{\ket}[1]{|#1\rangle}
\newcommand{\ketbra}[2]{|#1\rangle \langle #2|}
\newcommand{\kp}{\arccos(\frac{\omega - \epsilon}{2t})}
\newcommand{\ldos}{\mbox{L.D.O.S.}}
\renewcommand{\log}[1]{\mbox{log}\left(#1\right)}
\newcommand{\Log}{\mbox{log}}
\newcommand{\Modsq}[1]{\left| #1\right|^2}
\newcommand{\nb}{\textbf{Note: \ \ \ }}
\newcommand{\op}[1]{\hat {#1}}
\newcommand{\opket}[2]{\hat #1 | #2 \rangle}
\newcommand{\occ}{\mbox{Occ. Num.}}
\newcommand{\Real}[1]{\mbox{Re}\left(#1\right)}
\newcommand{\so}{\Rightarrow}
\newcommand{\sol}{\textbf{Solution: \ \ \ }}
\newcommand{\thetafn}[1]{\  \! \theta \left(#1\right)}
\newcommand{\tin}{\int_{-\infty}^{+\infty}\! \! \!\!\!\!\!}
\newcommand{\Tr}[1]{\mbox{Tr}\left(#1\right)}
\newcommand{\kb}{k_B}
\newcommand{\rad}{\mbox{ rad}}
\renewcommand\thesection{\arabic{section}}

\title{Origami-based spintronics in graphene}

\author{A. T. Costa}
\email{antc@if.uff.br}
\affiliation{Instituto de F\'isica, Universidade Federal Fluminense, 24210-346 Niter\'oi, RJ, Brazil}
\author{M. S. Ferreira}
\email{ferreirm@tcd.ie}
\affiliation{School of Physics and CRANN, Trinity College Dublin, Dublin 2, Ireland}
\author{Toby Hallam}
\affiliation{Centre for Research on Adaptive Nanostructures and Nanodevices (CRANN), Trinity College Dublin, Dublin 2, Ireland}
\author{Georg S. Duesberg}
\affiliation{School of Chemistry and CRANN, Trinity College Dublin, Dublin 2, Ireland}
\author{A. H. Castro Neto}
\affiliation{Graphene Research Centre and Department of Physics, National University of Singapore, 6 Science Drive 2, 117546, Singapore}
\affiliation{Department of Physics, Boston University, 590 Commonwealth Ave., Boston, MA, 02215, USA}

\date{\today}

\begin{abstract}
We show that periodically folded graphene sheets with enhanced spin-orbit interaction due to curvature effects can carry spin polarized currents and have gaps in the electronic spectrum in the presence of weak magnetic fields. Our results indicate that such origami-like structures can be used efficiently in spintronic applications. 
\end{abstract}


\maketitle
\bibliographystyle{abbrv} 

\section{Introduction}
Often hailed as a wonder material due to its impressive physical properties \cite{geim_review}, graphene has opened several venues of basic science exploration and it is a material that has a tremendous technological potential. Despite its potential applicability, the lack of a bandgap is a well known limitation that currently prevents the use of graphene in digital electronic applications \cite{roadmap}. Different strategies have been attempted to remedy this shortcoming, namely, by quantum confinement in the form of nanoribbons and quantum dots \cite{ribbon,ribbon2}, by stacking graphene sheets in bilayers in the presence of a perpendicular electric field \cite{bilayer,bilayer2}, by strain-engineering its electronic structure \cite{strain-engineering, s-e2, s-e3}, or by simply chemically doping the graphene sheets \cite{chemical,chemical2}. Unfortunately, these attempts have so far failed to produce technologically relevant semiconducting graphene due to several difficulties that go from the small size of the gaps they produce to the disorder that they introduce. 

On a different front, the field of spintronics appears as one of the most promising areas for graphene since the extremely small spin-orbit interaction (SOI) of carbon makes the spin dissipation that otherwise exists in most materials practically negligible \cite{SO1, SO2}. This suggests that information stored in the electronic spin of graphene can be retained for times considerably longer than in ordinary metals. Furthermore, this information can  travel  longer distances with very little loss \cite{waveguide, pumping, van}. Not surprisingly, there is a growing interest in graphene-based spintronics as demonstrated by the volume of recent literature on the topic \cite{macdonald}.  

Driven by the necessity of a bandgap and by the growing interest in graphene-based spintronics, in this manuscript we propose a simple mechanism that not only produces a gapped electronic structure in graphene but that also spin-polarizes its current. We show that this effect arises quite simply by the combined presence of two key ingredients: the SOI and an externally applied magnetic field. While magnetic fields are controllable, the SOI of a material is normally constant and small in the case of carbon. Therefore, it might seem too ambitious to amplify both ingredients enough for the appearence of a possible gap. Nevertheless, recent  discoveries have demonstrated that the SOI is enhanced when graphene is mechanically bent away from its planar geometry \cite{s-o-enhanced1,s-o-enhanced3,kawakami,barbaros} suggesting that folding might function as a viable mechanism to induce a bandgap. In fact, here we show that folded graphene sheets in the presence of externally applied magnetic fields may display both a bandgap and spin-polarized currents. Not possible with bulk 3-dimensional structures, folding may pave the way to a whole new approach of dealing with spin electronics in 2-dimensional systems, giving rise to the so-called origami spintronics.  

\section{Model and Results}
\begin{figure}[t]
\centering
\includegraphics[width =0.45\textwidth]{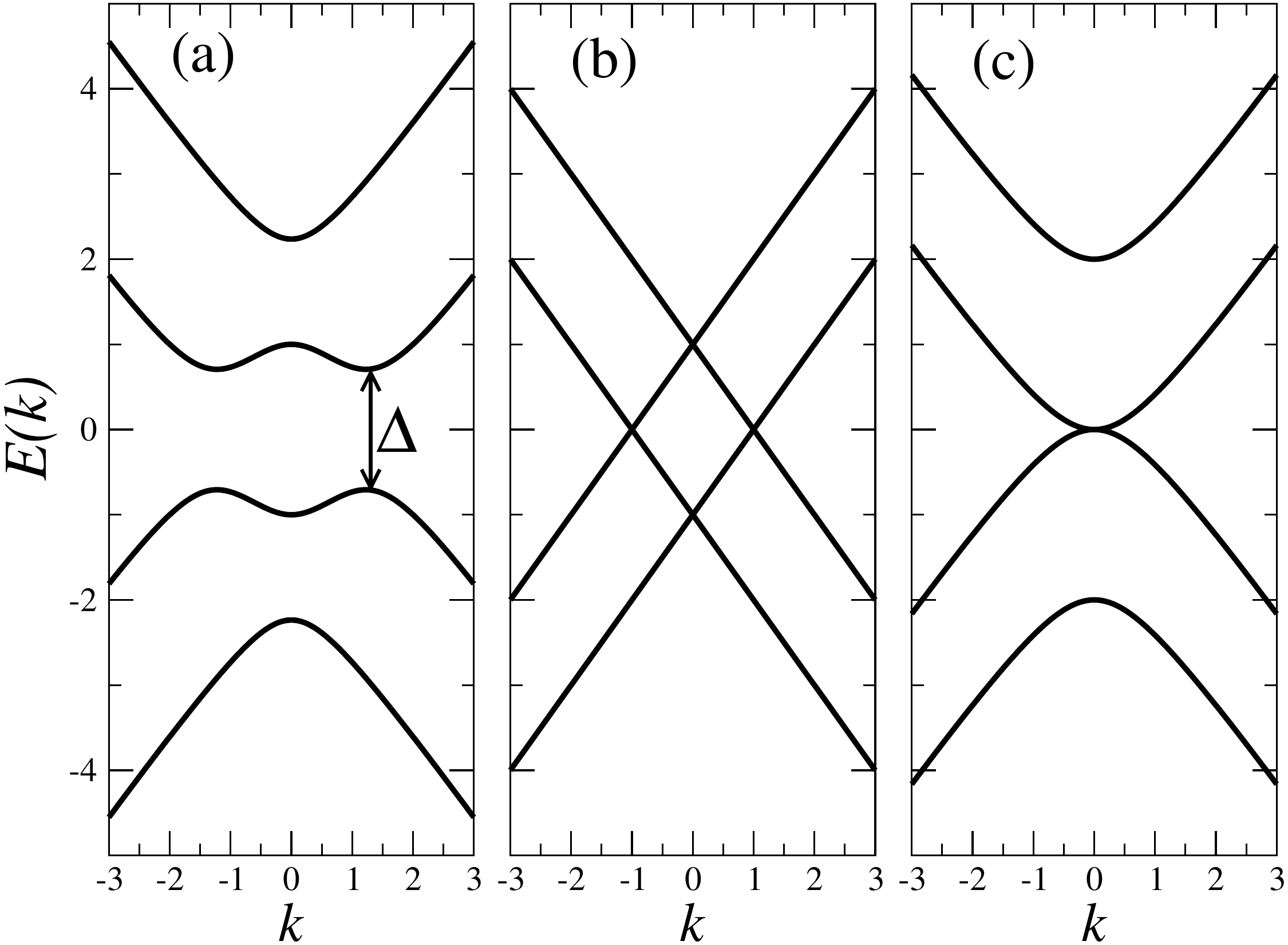}
\caption{Dispersion relation $E(k)$ obtained from the four eingevalues of the Hamiltonian $H$. (a) In general the electronic structure displays a distinctive bandgap $\Delta$, depicted by the vertical arrow. The energy gap disappears when (b) $\gamma\neq0$; $\lambda=0$ or (c) $\gamma=0$; $\lambda\neq0$ }
\label{fig1}
\end{figure}

It is convenient to start by showing that the two aforementioned key ingredients indeed lead to a bandgap in graphene. Let us  consider a graphene sheet with SOI and in the presence of an 
externally applied magnetic field. For energies close to the Dirac point, the band Hamiltonian $H$ is commonly described by a 2-dimensional Dirac equation. 
In this case, the Hamiltonian is given by \cite{KaneMele2005}: ${\hat H} = v_F {\hat 1} \otimes {\vec \sigma} \cdot {\vec p}+2\lambda(\sigma_x \otimes \tau_y + \sigma_y \otimes \tau_x)
+\gamma {\hat 1}\otimes \tau_z$, where $v_F$ is the Fermi velocity, $\sigma_{x,y,z}$ are Pauli matrices acting on the sublattice space, $\tau_{x,y,z}$ are Pauli 
matrices acting on the spin subspace, ${\vec p}= - i \hbar {\vec \nabla}$  is the momentum operator, $\lambda$ represents the strength of the SOI and $\gamma$ 
is the Zeeman-field factor. Note that the Hamiltonian recovers the standard form for pristine graphene when both $\gamma$ and $\lambda$ vanish, so it is convenient 
to express the Hamiltonian ${\hat H} = {\hat H}_0 + {\hat V}$, where ${\hat H}_0$ is the pristine Hamiltonian and ${\hat V}$ accounts for the $\gamma$- and 
$\lambda$-dependent perturbation. When expressed on the basis formed by the vector $\Psi = (\psi_a^\uparrow,\psi_a^\downarrow,\psi_b^\uparrow,\psi_b^\downarrow)^T$, 
where $\psi_\zeta^\sigma$ is the electron wavefunction on sublattice $\zeta$ with spin $\sigma$, the Hamiltonian is written in matrix form as:
\begin{equation}
H(k_x,k_y) = \begin{pmatrix}
-\gamma & 0 & k_x + i k_y & -2 i \lambda\\
0 & \gamma & 0 & k_x + i k_y \\
k_x - i k_y & 0 & -\gamma & 0 \\
2 i \lambda & k_x - i k_y & 0 & \gamma
\end{pmatrix}
\label{matriz}
\end{equation}
The four corresponding eigenvalues are:
\begin{equation}
E(k)= \pm \sqrt{k^2 + \gamma^2 + 2 \lambda^2 \pm 2 \sqrt{\lambda^4 + k^2 (\gamma^2 + \lambda^2)}}\,\,,
\label{dispersion}
\end{equation}
where $k^2 = k_x^2 + k_y^2$. Figure~\ref{fig1}(a) shows the dispersion $E(k)$ for arbitrary values of $\lambda$ and $\gamma$. Note the distinctive energy gap 
$\Delta = {2 \lambda \over \sqrt{1 + (\lambda / \gamma)^2}}$ around $E=0$. The characteristic linear dispersion relation of pristine graphene reappears 
from Eq. (\ref{dispersion}) when $\gamma=\lambda=0$. It is worth pointing out that either $\gamma$ or $\lambda$ alone is not capable to induce the 
bandgap opening, as shown in Figs.~\ref{fig1}(b) and \ref{fig1}(c). Only when both quantities are non-zero, will the bandgap appear. This is explained by the fact 
that $\gamma$ and $\lambda$ define two complementary energy scales and the magnitude of the bandgap $\Delta$ is determined ultimately by the 
smallest of these two quantities. 

\begin{figure}[t]
\centering
\includegraphics[width =0.45\textwidth]{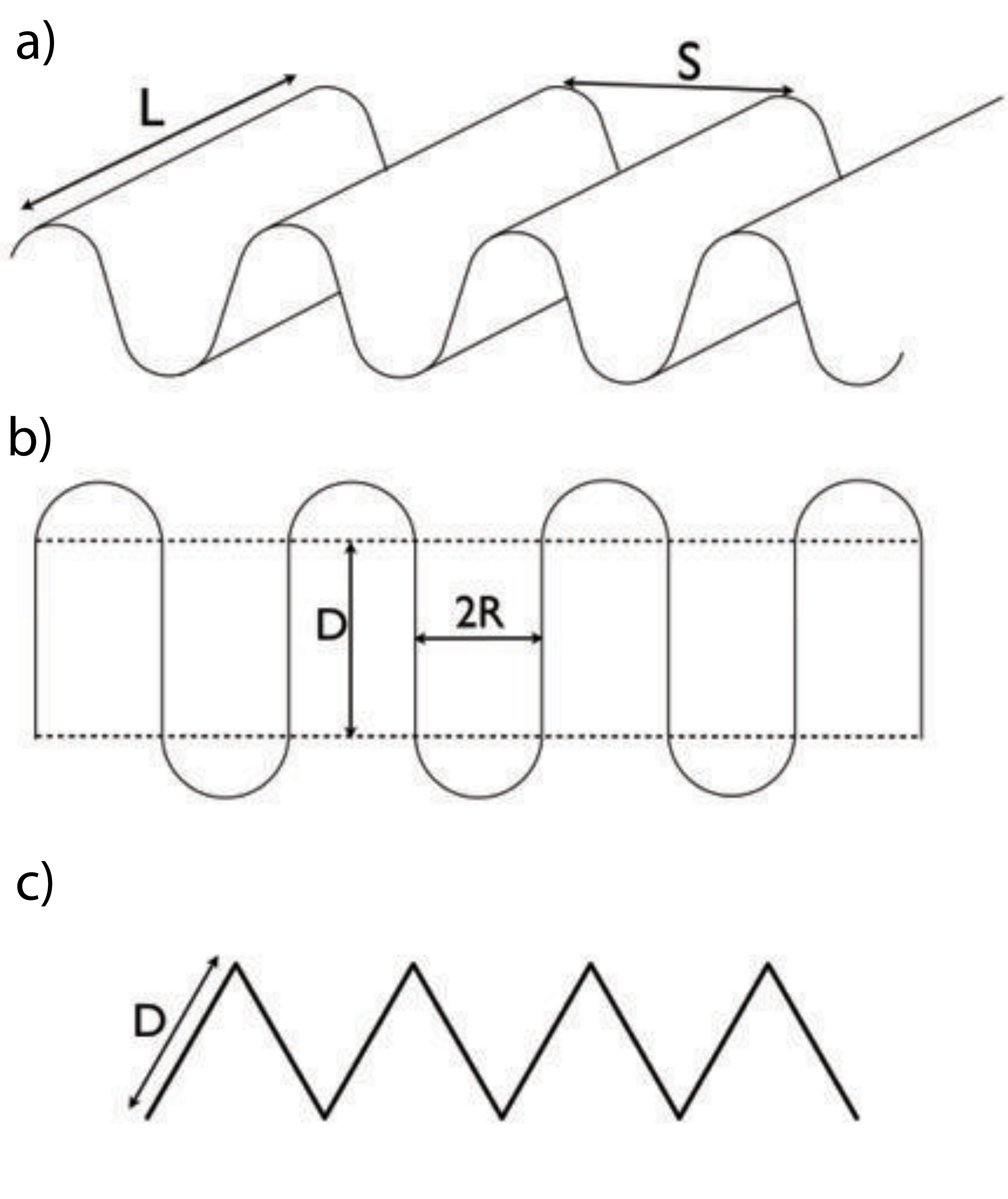}
\caption{Schematic diagrams of the corrugated geometry adopted by our graphene sheets. (a) Characteristic shape of corrugated sheets of graphene of width $L$ with a periodicity $S$. For practical purposes we assume $L \rightarrow \infty$. (b) Cross section of the corrugated graphene sheet used in our calculations, whereby sections of curved graphene with curvature radius $R$ and width $\pi R$ are spaced by flat regions of width $D$. For distinction, these two different regions are separated by the horizontal dashed lines. (c) In the limit $R \rightarrow 0$, the corrugated sheet becomes a series of equally spaced sharp creases separated by flat sections of graphene. }
\label{fig2}
\end{figure}

Let us now consider folding the graphene sheet as a SOI-enhancing mechanism. In this case, however, the SOI is not homogeneously distributed but spatially limited to the region of the graphene sheet surrounding the non-planar deformations. We consider a corrugated structure where curved deformations to the graphene sheet are ordered in a periodic structure. This can be achieved by patterning a substrate in the form of periodic trenches \cite{experimental-array,duesberg} on top of which a graphene sheet is deposited. In this way, the graphene sheet alternates between curved and flat regions, the dimensions of which will depend on the geometry of the patterned substrate. Figure~\ref{fig2}(a) shows an image of a typical corrugated sheet composed of a sequence of parallel ridges and troughs. While the diagram depicts a sheet of width $L$, 
it is supposed to represent a system where $L \gg S$, meaning that for practical purposes $L \rightarrow \infty$. For simplicity, we consider the corrugated graphene sheet made of a series of curved regions, all composed of semi-circular cross sections with curvature radius $R$, equally spaced by a distance $D$, the cross section of which is schematically shown in Fig.~\ref{fig2}b. This setup is equivalent to an array of half nanotubes joined seamlessly together by flat nanoribbons. Since the material has to comply during folding, this kind of distortion of the graphene sheet does not introduce the disorder that is so deleterious to the transport properties in other gap-opening proposals. 

Differently from the previous case where we looked at how the electronic band structure changes with the inclusion of the SOI plus the Zeeman term across the entire system, in the corrugated case 
we investigate how the conductance of the material responds to the addition of exactly the same ingredients, now in a periodic arrangement as shown in Fig.~\ref{fig2}b. In this case both these ingredients are added to the pristine Hamiltonian only where the graphene sheet has a non-zero curvature. The advantage of studying the transport response of the system is that we are able not only to identify the opening of a bandgap, which appears as a region of zero conductance, but we can simultaneously investigate the spin dependence of the transport response when the conductance is finite. In this way, we can study the appearance of a bandgap and assess its potential impact on spintronics applications at the same time. 

Rather than using the Dirac equation to describe the electronic structure of the system, we consider the tight-binding (TB) Hamiltonian whose accuracy is not limited to a narrow energy range surrounding the Dirac point \cite{beyond-linear}.  The Hamiltonian is written as ${\hat H} = \sum ({\hat a}^\dagger_{j,\sigma} \,{\hat b}_{j^\prime,\sigma}+\mathrm{h.c.}) + {\hat V}$, where the operators ${\hat a}^\dagger_{j,\sigma}$ (${\hat a}_{j,\sigma}$) creates (anihilates) an electron at site $j$ and spin $\sigma$ on the sublattice $A$. The ${\hat b}^\dagger_{j,\sigma}$ and ${\hat b}_{j,\sigma}$ are the corresponding operators for the B sublattice. The operator ${\hat V} = i\lambda\sum_{ll'}\sum_{\alpha\beta}\hat{z}\cdot(\vec{\sigma}_{\alpha\beta}\times\vec{d}_{ll'})c^\dagger_{l\alpha}c_{l'\beta} - \gamma\sum_{l\alpha}\alpha c^\dagger_{l\alpha}c_{l\alpha}$ represents the effects of a Rashba spin-orbit coupling \cite{KaneMele2005} plus a Zeeman field. $\vec{d}_{ll'}$ is the position of site $l'$ relative to site $l$. $\vec{\sigma}$ is the vector formed by the Pauli matrices and the index $\alpha$ in the last sum takes the value +1 for majority spins and -1 for minority spins. The Rashba term arises whenever inversion symmetry in the direction perpendicular to the graphene sheet is broken, a condition realized for samples deposited on a substrate or in the presence of an electric field perpendicular to the sheet plane.

The parameters $\lambda$ and $\gamma$ are dependent on the curvature and on the applied magnetic field, respectively. The former can be estimated from References \cite{SO1, SO2}, 
which studied the SOI in carbon nanotubes, and the latter can be extracted from characteristic values of magnetic fields. For the case shown in Fig.~\ref{fig2}b the potential 
${\hat V}$ acts only where the curved regions are located. In other words, we must consider a graphene  sheet with an array of M equally spaced regions of width $\pi R$ separated by a distance $D$. Following Refs. \cite{SO1, SO2}, $\lambda$ is inversely proportional to the radius $R$, which means that higher curvatures 
correspond to narrower curved sections with correspondingly stronger values for $\lambda$. In the extreme limit of very high curvatures ($R \rightarrow 0$) the width of the regions with non-zero SOI becomes small, giving rise to a geometry depicted by Fig.~\ref{fig2}c. In all cases, we assume that the corrugated structure is connected by leads made of semi-infinite pristine graphene. 
 
The conductance is calculated by the Kubo formula expressed in terms of the single particle Green function $G$, which in turn, is defined as 
$G(E) = (E\, {\hat 1} - {\hat H})^{-1}$, where $E$ is the energy and ${\hat 1}$ is the identity operator. The zero bias DC conductance $\Gamma$ is written as: 
\begin{equation}
\Gamma = \frac{4e^2}{h}\mathrm{Re}\mathrm{Tr}\left( [\tilde{G}]_{00}t_{01}[\tilde{G}]_{11}t_{10}
-t_{01}[\tilde{G}]_{10}t_{01}[\tilde{G}]_{10}.
\right)
\label{conduc}
\end{equation}
Here, $\tilde{G}_{l\alpha,l'\beta} = [G_{l\alpha,l'\beta}-(G_{l'\beta,l\alpha})^*]/2i$ and $[G]_{AB}$ 
is the matrix formed by the Green function elements connecting unit cells $A$ and $B$. $\alpha$ and $\beta$ are spin indices.  We define unit cells as lines along the ``armchair'' direction. The trace is taken over both site and spin indices. All Green functions above are evaluated at the Fermi level.

\begin{figure}[t]
\begin{tabular}{cc}
\begin{minipage}{0.5\textwidth}
\includegraphics[width=0.96\textwidth]{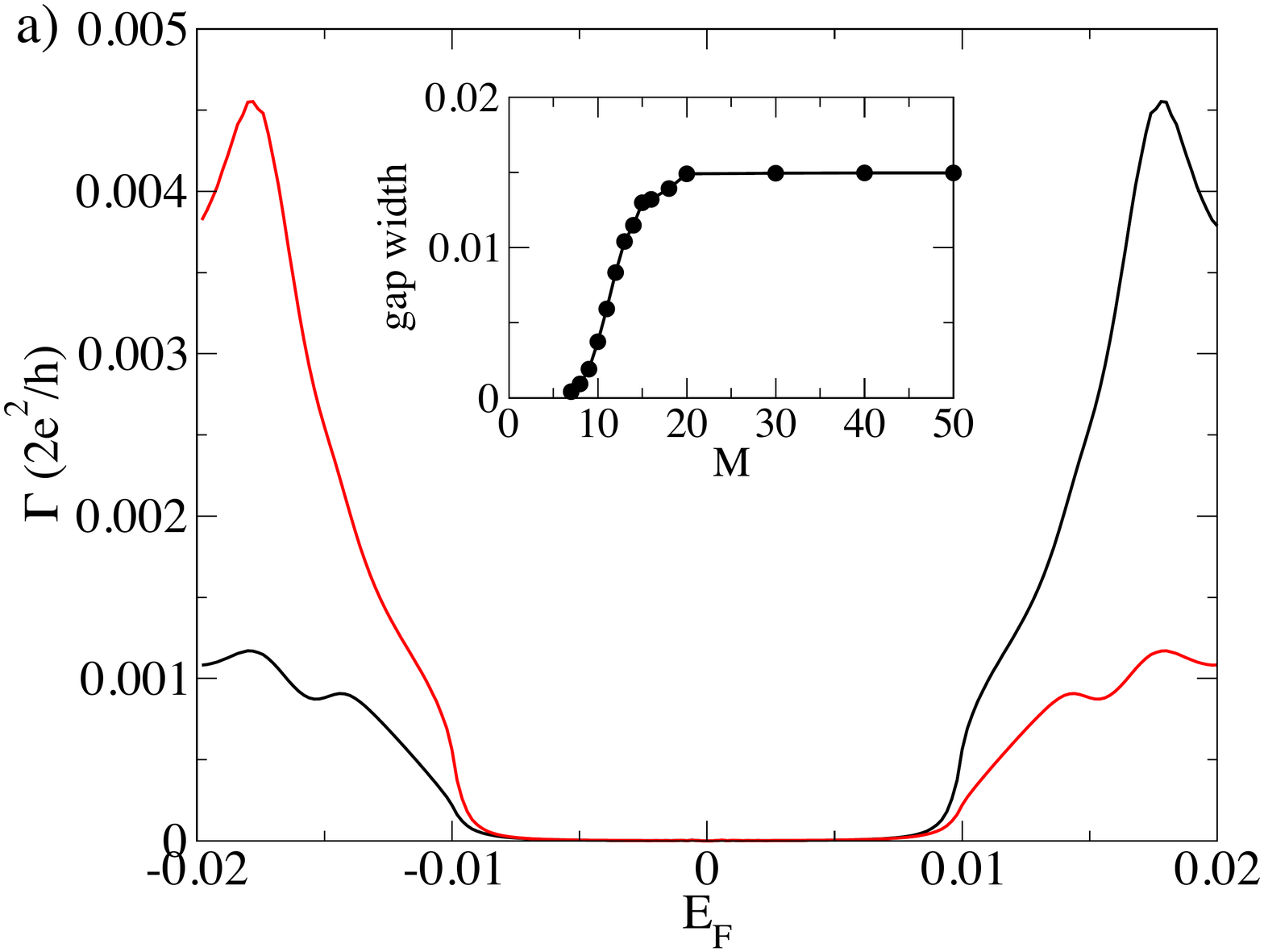}
\end{minipage} &
\begin{minipage}{0.5\textwidth}
\includegraphics[width=0.96\textwidth]{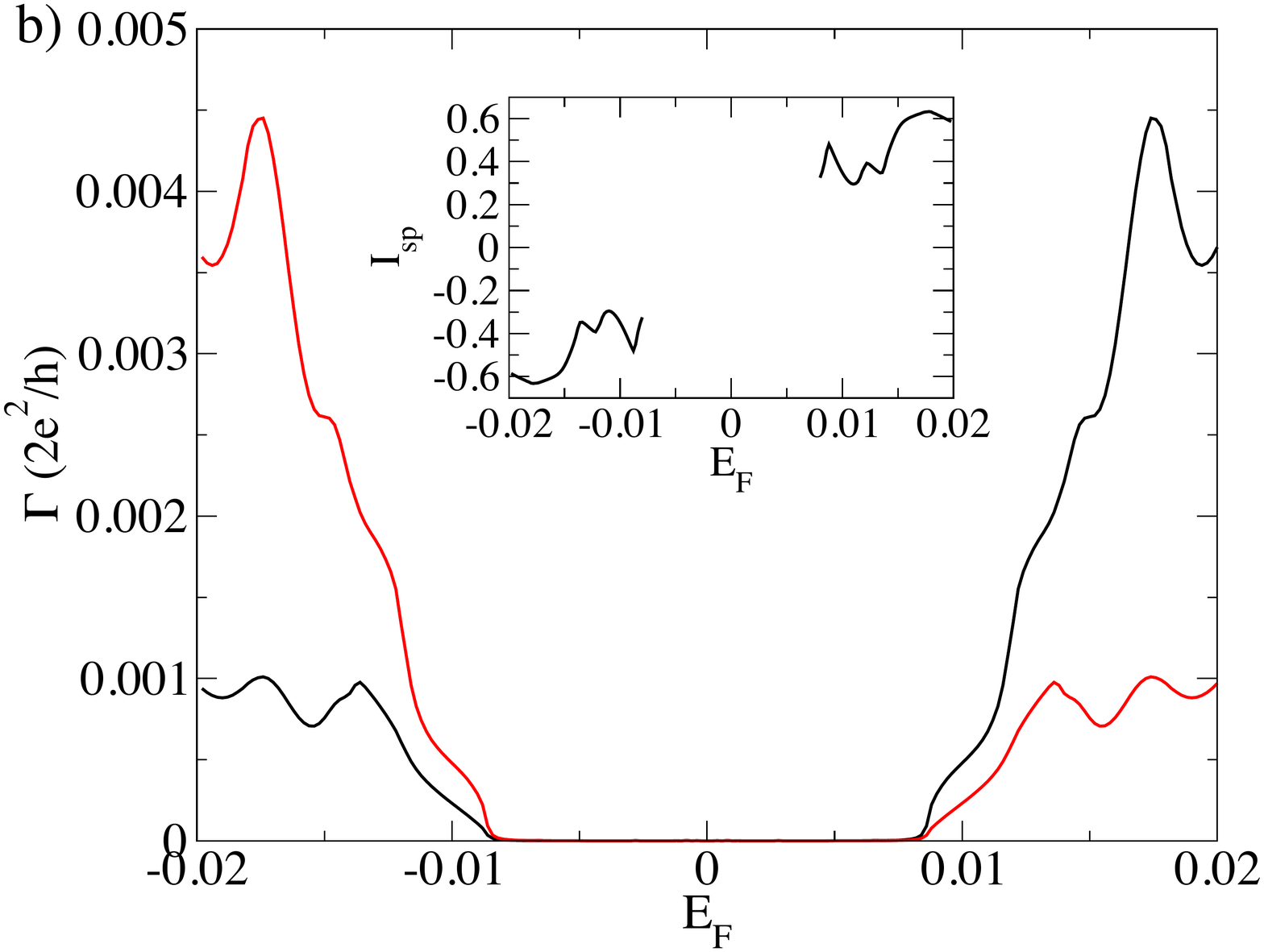}
\end{minipage} \\
\begin{minipage}{0.5\textwidth}
\includegraphics[width=0.96\textwidth]{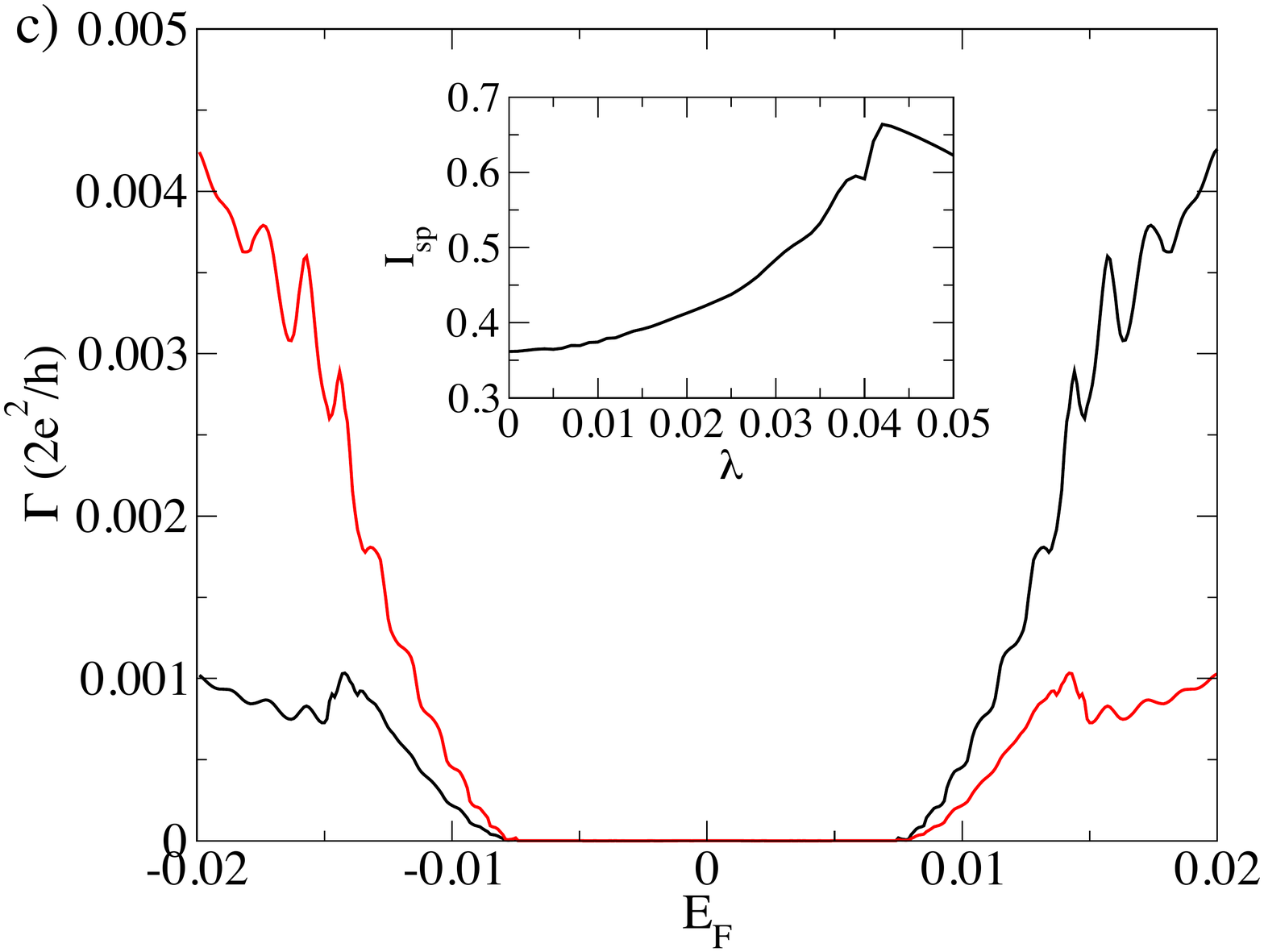}
\end{minipage} &
\begin{minipage}{0.5\textwidth}
\includegraphics[width=0.96\textwidth]{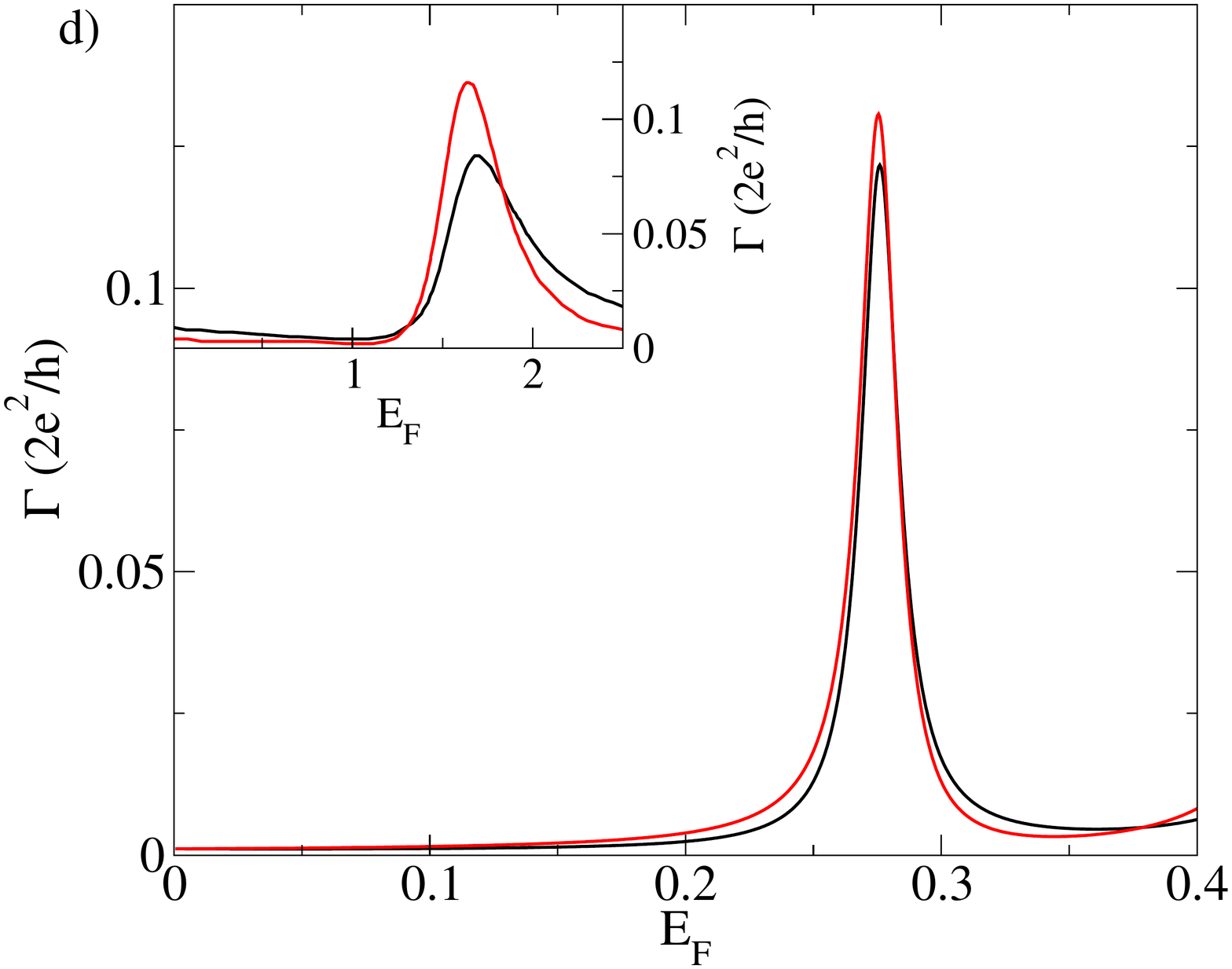}
\end{minipage} 
\end{tabular}
\caption{Spin-dependent conductances of a corrugated sheet made of $M$ curved regions of radius $R\sim 6$\AA\ separated by flat sections of graphene of width $D = 20$. Except for the insets, black (red) lines represent $\uparrow$ ($\downarrow$) spin-polarized conductances. (a) $M = 10$; $\gamma = 5\times 10^{-2}$; $\lambda =5\times 10^{-2}$. The inset shows how the conductance gap scales with $M$. Note that it tends to a saturation value when $M \gg 1$. (b) $M = 15$; $\gamma =5\times 10^{-2} $; $\lambda =5\times 10^{-2}$. The inset depicts the spin polarization of the current $I_{sp}$ as a function of $E_F$. Only results outside the bandgap are shown. (c) $M = 50$; $\gamma = 5\times 10^{-2}$; $\lambda = 5\times 10^{-2}$. Shown in the inset is the dependence of $I_{sp}$ on the SOI parameter $\lambda$. (d) Results for creased graphene sheets represented schematically by Fig.~\ref{fig2}(c) for $M=2$ delta functions for electrons constrained to move perpendicularly to the creases calculated based on the TB-model (main figure) and on the simplified model (inset).}
\label{fig3}
\end{figure}

In Fig.~\ref{fig3} we show the results for the spin-dependent conductance as a function of the Fermi level $E_F$ for $M =10, 15, \, {\rm and}\, 50$. Note the opening of a gap as $M$ increases, as shown in the inset of Fig.~\ref{fig3}(a). The magnitude of the gap increases with $M$ but saturates when $M \gg 1$. 
In addition, outside the gapped energy region, the conductance becomes spin dependent, {i.e.}, the transport properties for the $\uparrow$ and $\downarrow$ spin electrons become asymetric, which is a basic requirement for exploring the use of any material in spintronics. Such an asymmetry in the spin-dependent conductance means that these corrugated systems can function as spin polarizers for the electronic currents in graphene. The inset of Fig.~\ref{fig3}(b) shows the spin-polarization of the current, 
defined as $I_{sp} = (I_\uparrow - I_\downarrow)/(I_\uparrow + I_\downarrow)$, as a function of $E_F$ whereas the inset of Fig.~\ref{fig3}(c) depicts $I_{sp}$ as a function of $\lambda$. 
Note the sign change of $I_{sp}$ on either side of the band gap. This feature is seen in all our results. 

To assess the robustness and generality of these findings we make use of a simplified approach that is less computationally involved and yet capable of capturing the contribution from the two key ingredients, namely the SOI and the Zeeman term. We investigate the limit $R \rightarrow 0$, shown schematically in Fig.~\ref{fig2}(c). In this limit, the SOI-enhancing curvature becomes extremely large at the same time as the extension of the curved 
region becomes vanishingly small. The smooth ondulations shown in Figs.~\ref{fig2}(a) and~\ref{fig2}(b) give way to equally spaced creases on the graphene sheet, ideally represented by Dirac delta functions that mark the locations where both the SOI and the Zeeman term are non-zero. Moreover, a further simplification is to assume electrons moving along the direction normal to the creases. The Hamiltonian in this case becomes ${\hat H} = {\hat H}_0 + \sum_{j=1}^M {\hat V} \, \delta(x-j D)$, where the operators ${\hat H}_0$ and ${\hat V}$ have been previously defined, the only difference being that $k_y=0$ and $k=k_x$. In other words, the problem in question is now solved by considering Dirac equation for massless particles under the action of a perturbing potential made of a series of $\lambda$- and $\gamma$-dependent delta functions \cite{gomes}. It is now straightforward to calculate the transmission coefficient of electrons across this potential and plot it as a function of the energy, which is proportional 
to the wavevector $k$. Since the conductance is also proportional to the transmission coefficient, we can now compare the results of this simplified approach with the TB-based results. In Fig.~\ref{fig3}(d) we show results for $M=2$ delta functions calculated with the simplified approach as well as with the TB adapted for normal incidence. Quantitative agreement is not expected because the discrete delta function representation in the TB-model is fundamentally different from the continuous one. From a qualitative point of view, however, both methodologies lead to the appearance of a low-energy conductance gap followed by a spin-polarized current for energies outside the bandgap. 
This commonality found with two distinct methodologies is indicative of the robustness of our findings, which is further confirmed by recent calculations based on Density Functional Theory results also displaying bandgap features when the SOI is enhanced by curvature effects \cite{claudia}. 

\section{Experimental feasibility}
In addition to folding, adsorbants that deform the pristine planar structure of graphene sheets also function as SOI-enhancers. Therefore, it is possible to combine them both such that SOI-enhancing adsorbants might get attached to mechanically curved regions which themselves possess enhanced spin-orbit coupling. Moreover, if the adsorbants have magnetic moments their exchange splitting may induce a spin splitting within
the underlying carbon atoms via hybridisation. The induced splitting will be equivalent to a large Zeeman field. While we are not able to quantify the precise contributions coming from each one of the different mechanisms, curvature and dopants acting jointly together will certainly amplify the overall effect of spin-polarization of the charge currents as well as the bandgap in the electronic structure of the system. 

Regarding the experimental feasibility of the system considered in this theoretical study, the system schematically depicted by Fig. 2(a) is easily realized if graphene sheets are deposited on periodically trenched substrates\cite{duesberg}. Despite being experimentally more challenging, the creased structures shown in Fig. 2(c) are also possible and are currently grown with impressive levels of control on their  dimensions and periodicity as illustrated in Figure 4. The graphene is transferred via a PMMA support layer onto a trenched PDMS stamp with a conventional polymer assisted process\cite{Li-Science}.The PMMA is then removed with a solvent bath and with controlled drying conditions the graphene film sinks into the trenches on the stamp. Upon further drying of the graphene, it partly adheres to the trench walls with a suspended configuration over the middle of the trench.  Printing the graphene inked stamps leads to the suspended graphene collapsing onto the substrate and the adhered side-wall graphene folding into a bi-layer configuration as shown in Figure 4a. As a result, freestanding fin-like structures are formed\cite{duesberg-small}. A scanning electron micrograph is shown in Figure 4b and an AFM image of a small portion of such a printed surface is depicted in Figure 4c. It is worth emphasizing that this methodology gives us full control on the dimensions and periodicity of the folded fins. The remaining challenge consists of carrying out transport measurements across such structures but the fact that they are grown with such an excellent level of control is the first step towards the experimental confirmation of our theoretical predictions

\section{Conclusions}
In summary, we have shown that by adding the Zeeman term and the SOI to the Hamiltonian of pristine graphene, we force the opening of a bandgap in its electronic structure and induce its charge current to become spin polarized. Both additions to the Hamiltonian are experimentally feasible and are mimicked by an externally applied magnetic field and by deforming the graphene sheet out of its planar geometry through folding. With the increasing degree of experimental control on both the chemistry and the geometry of nanoscaled surfaces, it is possible that by engineering the shape and composition 
of graphene sheets we can create spin-polarized currents in an energy-gapped material. Such an origami-like technique is likely to bring about a whole new range of spintronic features in 2-dimensional systems not possible in 3-dimensional structures.

\begin{acknowledgments}
MSF acknowledges financial support from Science Foundation Ireland under Grant Number SFI 11/RFP.1/MTR/3083.
AHCN acknowledges DOE grant DE-FG02-08ER46512, ONR grant MURI N00014-09-1-1063, and the NRF-CRP award "Novel 
2D materials with tailored properties: beyond graphene" (R-144-000-295-281). A. T. C. acknowledges partial financial 
support from CNPq (Brazil) and INCT Nanomateriais de Carbono. GSD acknowledges support from SFI under grant PI-10/IN.1/I3030. 
\end{acknowledgments}

\begin{figure}[t]
\centering
\includegraphics[width =0.55\textwidth]{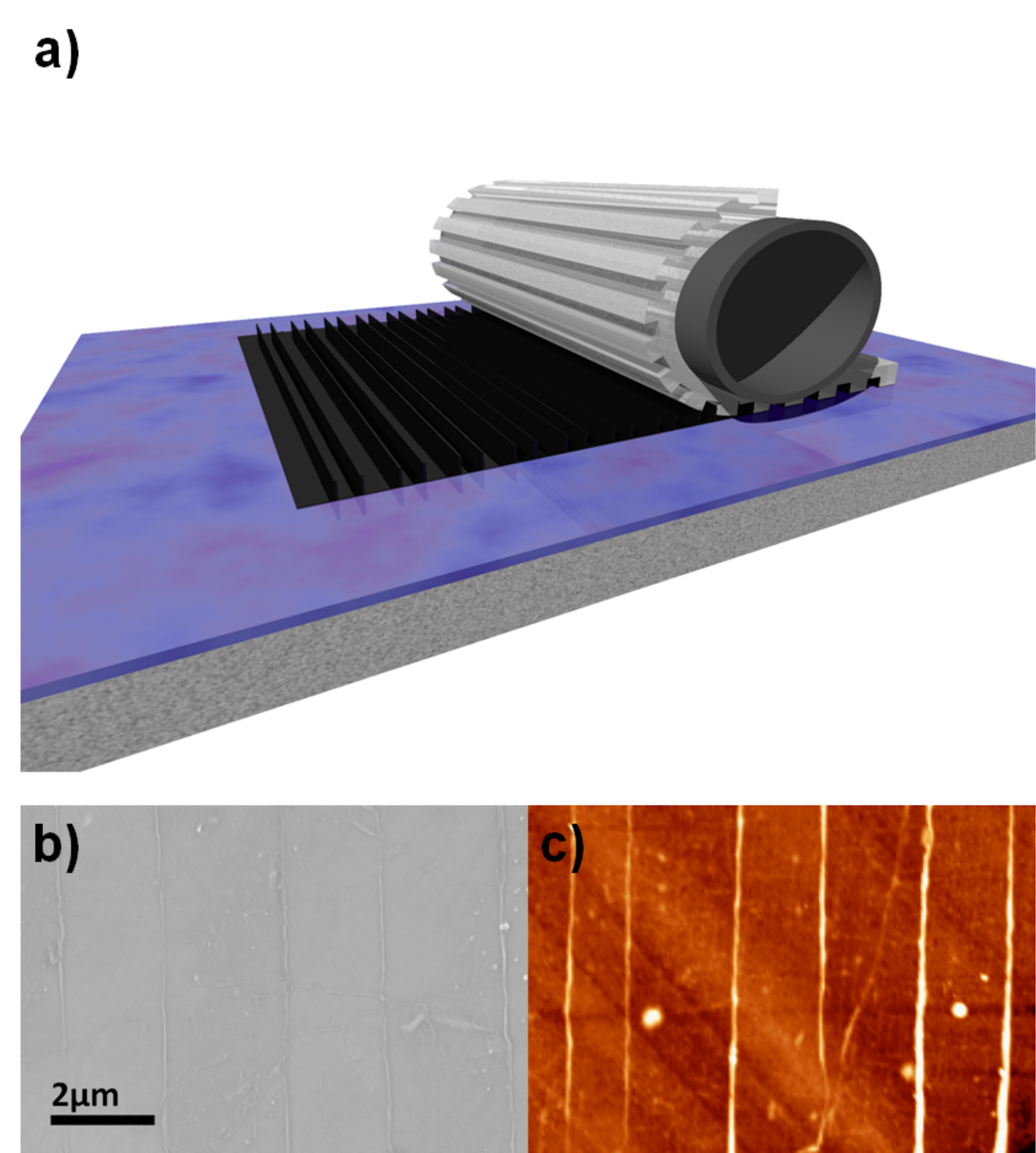}
\caption{Experimental method used to grow fin-like structures made of a single folded graphene sheet. (a) A graphene sheet draped over a relief patterned PDMS stamp is used to print free standing graphene fins in a modified transfer printing technique. (b) A scanning electron micrograph and (c) an AFM image of a small portion of a folded graphene surface.}
\label{fig4}
\end{figure}


\begin{thebibliography}{99}

\bibitem{geim_review} A. K. Geim, K. S. Novoselov, Nature Materials \textbf{6}, 183-191 (2007)
\bibitem{roadmap} K. S. Novoselov, V. I. Falko, L. Colombo, P. R. Gellert, M. G. Schwab and K. Kim, Nature, {\bf 490} 192-200 (2012)
\bibitem{ribbon} M. Han, B. Özyilmaz, Y. Zhang and Philip Kim, Phys. Rev. Lett. {\bf 98}, 206805 (4pp) (2007)
\bibitem{ribbon2} J. M. Cai, P. Ruffieux, R. Jaafar, M. Bieri, T. Braun, S. Blankenburg et al, Nature {\bf 466}, 470-473 (2010)
\bibitem{bilayer} T. Ohta, A. Bostwick, T. Seyller, K. Horn, E. Rotenberg, Science {\bf 313}, 951-954 (2006)
\bibitem{bilayer2} J. B. Oostinga, H. B. Heersche, X. Liu, A. F. Morpurgo and L. M. K. Vandersypen, Nature Mater. {\bf 7}, 151-157 (2008)
\bibitem{strain-engineering} V. M. Pereira, A. H. Castro Neto, and N. M. R. Peres, Phys. Rev. B {\bf 80}, 045401 (8pp) (2009)
\bibitem{s-e2} Z. H. Ni, T. Yu, Y. H. Lu, Y. Y. Wang, Y. P. Feng, and Z. X. Shen, ACS Nano {\bf 2}, 2301-2305 (2008)
\bibitem{s-e3} T. M. G. Mohiuddin, A. Lombardo, R. R. Nair, A. Bonetti, G. Savini, R. Jalil et al, Phys. Rev. B {\bf 79}, 205433 (8pp) (2009).
\bibitem{chemical} D. C. Elias, R. R. Nair, T. M. G. Mohiuddin, S. V. Morozov, P. Blake, M. P. Halsall et al, Science {\bf 323}, 610-613 (2009) 
\bibitem{chemical2} K. P. Loh, Q. Bao, P. K. Anga and J. Yanga, J. Mater, Chem. {\bf 20}, 2277-2289 (2010) 
\bibitem{SO1} D. Huertas-Hernando, F. Guinea and A. Brataas, Phys. Rev. B {\bf 74}, 155426 (15pp) (2006)
\bibitem{SO2} D. Huertas-Hernando, F. Guinea, and A. Brataas, Phys. Rev. Lett. {\bf 103}, 146801 (4pp) (2009) 
\bibitem{waveguide} F. S. M. Guimaraes, D. F. Kirwan, A. T. Costa, R. B. Muniz, D. L. Mills and M. S. Ferreira, Phys. Rev. B {\bf 81}, 153408 (4pp) (2010)
\bibitem{pumping} F. S. M. Guimaraes, A. T. Costa, R. B. Muniz and M. S. Ferreira, Phys. Rev. B {\bf 81}, 233402 (4pp) (2010)
\bibitem{van} N. Tombros, C. Jozsa, M. Popinciuc, H. T. Jonkman, and B. J. van Wees, Nature \textbf{448}, 571-574 (2007). 
\bibitem{macdonald} D. Pesin and A. H. MacDonald, Nature Mater. {\bf 11}, 409-416 (2012) and references therein 
\bibitem{kawakami} W. Han, J. Chen, D. Wang, K. M. McCreary, H. Wen, A. G. Swartz, J. Shi, and R. K. Kawakami, Nano Lett. \textbf{12}, 3443-3447 (2012); K. M. McCreary, A. G. Swartz, W. Han, J. Fabian, and R. K. Kawakami, Phys. Rev. Lett. \textbf{109}, 186604 (4pp) (2012).
\bibitem{barbaros} J. Balakrishnan, G. K. W. Koon, M. Jaiswal, A. H. Castro Neto, and Barbaros \"Ozyilmaz, Nature Physics \textbf{9}, 284–287 (2013) .
\bibitem{s-o-enhanced1} A. H. Castro Neto and F. Guinea, Phys. Rev. Lett. {\bf 103}, 026804 (4pp) (2009)
\bibitem{s-o-enhanced3} D. Ma, Z. Li and Z. Yang, Carbon {\bf 50}, 297-305 (2012)
\bibitem{KaneMele2005} C. L. Kane and E. J. Mele, Phys. Rev. Lett. \textbf{95}, 226801 (4pp) (2005)
\bibitem{so-measure1} Y. S. Dedkov, M. Fonin, U. Rudiger and C. Laubschat, Phys. Rev. Lett. {\bf 100}, 107602 (4pp) (2008)
\bibitem{so-measure2} O. Rader, A. Varykhalov,  J. Sánchez-Barriga, D. Marchenko, A. Rybkin and A. M. Shikin, Phys. Rev. Lett. {\bf 102} 057602 (4pp) (2009)
\bibitem{experimental-array} J. Hicks, A. Tejeda, A. Taleb-Ibrahimi, M. S. Nevius, F. Wang, K. Shepperd et al, Nature Phys. {\bf 9}, 49-54 (2013)
\bibitem{duesberg} S. Winters, T. Hallam, H. Nolan, and G. S. Duesberg, Phys. Status Solidi B \textbf{249}, 2515–2518 (2012)  
\bibitem{Li-Science} X. Li, W. Cai, J. An, S. Kim, J. Nah, D. Yang et al, Science {\bf 324}, 1312-1314 (2009)
\bibitem{duesberg-small} T. Hallam, M. T. Cole, W. I. Milne and G. S. Duesberg, Accepted for publication in Small (2013) 
\bibitem{beyond-linear} S. R. Power and M. S. Ferreira, Phys. Rev. B {\bf 83}, 155432 (7pp) (2011) 
\bibitem{gomes} J. V. Gomes and N. M. R. Peres, J. Phys. Condens. Matter {\bf 20}, 325221 (12pp) (2008)
\bibitem{claudia} S. M. Avdoshenko, P. Koskinen, H. Sevincli, A. Popov, C. G. Rocha, Scientific Reports \textbf{3}, 1632 (6pp) (2013)
\end{thebibliography}
\end{document}